\documentclass[10pt,conference]{IEEEtran}

\usepackage{cite}
\usepackage{amsmath,amssymb,amsfonts}
\usepackage{algorithmic}
\usepackage{graphicx}
\usepackage{textcomp}
\usepackage{xcolor}
\usepackage{enumitem}
\usepackage{tikz}
\usepackage{multirow}
\usepackage{subcaption}
\usepackage{float}
\usepackage[hyphens]{url}
\usepackage[]{hyperref}

\def\BibTeX{{\rm B\kern-.05em{\sc i\kern-.025em b}\kern-.08em
    T\kern-.1667em\lower.7ex\hbox{E}\kern-.125emX}}
\begin{document}

\newcommand*{\dhanvi}{\textcolor{orange}}
\newcommand*{\gokul}{\textcolor{red}}
\newcommand*{\yuewen}{\textcolor{blue}}

\pdfpagewidth=8.5in
\pdfpageheight=11in


\pagenumbering{arabic}

\title{Scalable Clifford-Based Classical Initialization for the Quantum Approximate Optimization Algorithm}

\author{
\IEEEauthorblockN{Dhanvi Bharadwaj}
\IEEEauthorblockA{
University of Michigan\\
Ann Arbor, MI, USA\\
dhanvib@umich.edu
}
\and
\IEEEauthorblockN{Yuewen Hou}
\IEEEauthorblockA{
University of Michigan\\
Ann Arbor, MI, USA\\
isaachyw@umich.edu
}
\and
\IEEEauthorblockN{Guangyi Li}
\IEEEauthorblockA{
University of Maryland\\
College Park, MD, USA\\
guangyi@umd.edu
}
\and
\IEEEauthorblockN{Gokul Subramanian Ravi}
\IEEEauthorblockA{
University of Michigan\\
Ann Arbor, MI, USA\\
gsravi@umich.edu
}
}

\maketitle
\thispagestyle{plain}
\pagestyle{plain}


\begin{abstract}
Variational Quantum Algorithms (VQAs), such as the Quantum Approximate Optimization Algorithm (QAOA), offer a promising route to tackling optimization problems on near- and intermediate-term quantum devices. However, their performance critically depends on the choice of initial parameters, and the limited expressiveness of the QAOA ansatz makes identifying effective initializations both difficult and unscalable. To address this, we propose a framework, \underline{S}calable \underline{P}arameter \underline{I}nitialization for \underline{Q}AOA (SPIQ)\footnote{SPIQ (pronounced \textit{``spike''}) is open source~\cite{spiq_github}.}, that employs a relaxed QAOA ansatz to enable classical search over a set of Clifford-preparable quantum states that yield high-quality solutions. These states serve as superior QAOA initializations, driving rapid convergence while significantly reducing the quantum circuit evaluations needed to reach high-quality solutions and consequently lowering quantum-device cost.

We present a scalable, application-agnostic initialization framework that achieves an absolute accuracy improvement of up to 80\% over state-of-the-art initialization and reduces initial-state diversity by an average of up to $10,448\times$ across QUBO, PUBO, and PCBO problems spanning tens to hundreds of qubits. We also benchmark its performance on a wide range of problem formulations and instances derived from real-world datasets, demonstrating consistent and scalable improvements over existing methods. Furthermore, we introduce two complementary strategies for selecting high-quality Clifford points identified by our search procedure and using them to seed multi-start optimization, thereby enhancing exploration and improving solution quality.

\end{abstract}
\section{Introduction}

Quantum computers offer the promise of computational advantage, especially in key areas like chemistry~\cite{kandala2017hardware}, optimization~\cite{moll_optimization}, and machine learning~\cite{biamonte2017quantum}. 
Quantum computing is rapidly progressing beyond small-scale, Noisy Intermediate-Scale Quantum (NISQ) devices~\cite{preskill2018quantum} toward systems with thousands of qubits supported by sophisticated error mitigation and potentially lightweight (or partial) error correction techniques~\cite{Preskill_2025,google_roadmap,iroadmap_2,eft}. These emerging systems may enable practical utility~\cite{kim2023evidence} by executing circuits with a non-trivial number of logical qubits and depths on the order of thousands of gates~\cite{Preskill_2025}. However, they will still fall short of supporting long-duration quantum algorithms~\cite{Shor_1997,grover1996fast}, which require millions of qubits, full fault tolerance, and the ability to perform billions of quantum operations~\cite{O_Gorman_2017}.

\begin{figure}[t]
    \centering
    \includegraphics[width=\columnwidth,trim={0.8cm 9.5cm 7cm 3.5cm},clip]{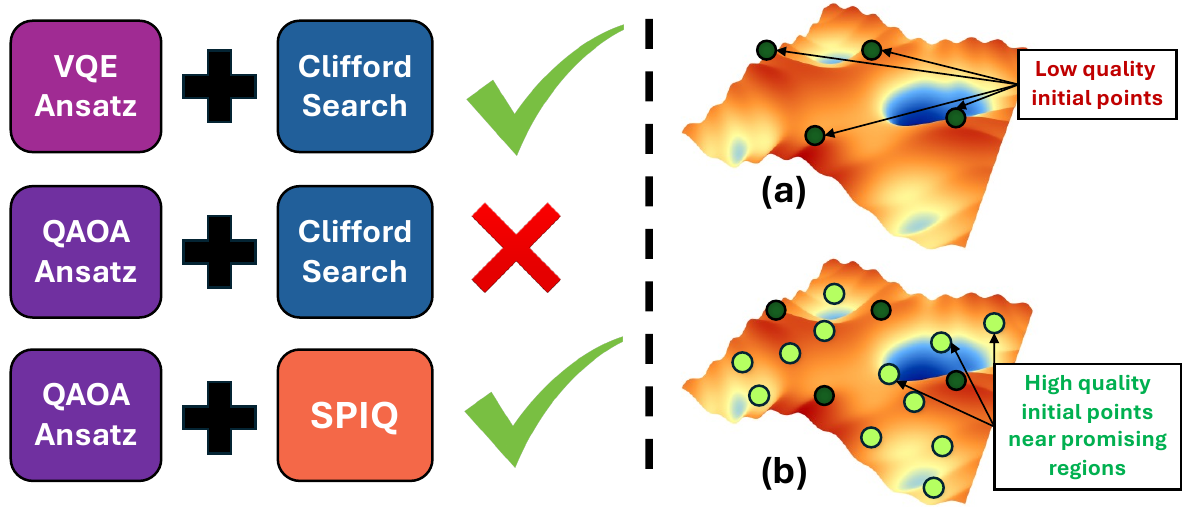}
    \caption{ (Left): High-level illustration of initialization frameworks for VQAs. The state-of-the-art Clifford approach (CAFQA) performs well for VQE but fails on standard QAOA due to its extremely limited optimization landscape. In contrast, our work (SPIQ) with QAOA yields robust and effective initialization.
    (Right): (a) Only a few, low-quality Clifford states are found with CAFQA, while (b) SPIQ finds many high-quality Clifford points.}
    \label{fig:intro_landscape}
\end{figure}

There are certain algorithms that exhibit an innate robustness to noise, allowing them to generate valuable results even without full fault tolerance. Among these, Variational Quantum Algorithms (VQAs)~\cite{cerezo21_vqa} are especially notable for their broad applicability. There is optimism within the community~\cite{zimborás2025mythsquantumcomputationfault} that we should continue to advance them in the hope of achieving real-world utility on next-generation quantum systems in areas such as combinatorial optimization~\cite{farhi14}, physics~\cite{kim2023evidence}, and chemistry~\cite{peruzzo2014variational}. VQAs are hybrid algorithms in which a parameterized quantum circuit is iteratively refined by a classical routine. Over successive optimization iterations, the VQA explores the solution landscape by minimizing the expectation value of the cost Hamiltonian with respect to the variational state.

Within the broader class of VQAs, one algorithm specifically designed to approximately solve combinatorial optimization problems is the Quantum Approximate Optimization Algorithm (QAOA) ~\cite{farhi14}.
QAOA alternates between applying problem-specific and mixing Hamiltonians, with the goal of steering the parameterized state toward a state that encodes a high-quality solution. The algorithm is parameterized by a set of a few angles that define the rotation gates, which are tuned by a classical optimizer to minimize the expectation value of the final quantum state. As the circuit depth $p$ increases, QAOA can, in principle, approximate the optimal solution with increasing accuracy, making it a promising candidate for near-term quantum advantage in discrete optimization tasks.

As with many optimization algorithms, the choice of initial parameters in QAOA strongly influences solution quality, and good initialization can substantially reduce quantum and classical resource costs by cutting the number of required circuit evaluations and improving the accuracy of the converged solution ~\cite{Sack_2021,blekos2024review}. However, searching for these initial parameters, especially in high-dimensional and noisy landscapes, is computationally intensive. This challenge is amplified when evaluations must be performed on actual quantum hardware, where each function call may require numerous circuit executions (shots), making the process resource- and time-consuming. Previous initialization strategies for QAOA have largely been tailored to specific problem instances \cite{wang2024red, INTERP, wilkie2024angle}, often relying on domain-specific knowledge or heuristics. Furthermore, many of these methods have typically been tested only on unconstrained or small-scale problems and on unweighted graph instances, and have not demonstrated effectiveness at larger scales.

To the best of our knowledge, there is currently no general-purpose initialization framework that is both broadly applicable and scalable across diverse problem types. These limitations pose a significant barrier to the widespread and efficient deployment of QAOA across the full spectrum of combinatorial optimization problems~\cite{lucas2014ising}, including Quadratic Unconstrained Binary Optimization (QUBO), Polynomial Unconstrained Binary Optimization (PUBO), and Polynomial Constrained Binary Optimization (PCBO).

To find a promising initialization strategy for the QAOA, we look to the initialization landscape of another VQA, the Variational Quantum Eigensolver (VQE)~\cite{peruzzo2014variational}. Among many techniques, VQE has particularly benefited from scalable and classically simulable strategies based on Clifford states~\cite{CAFQA_Ravi2022,wang2024demonstrationcafqabootstrappedvariationalquantum}. That framework, CAFQA, demonstrates strong performance on quantum chemistry tasks and highlights the value of high-quality initializations for improving convergence and solution quality. However, directly applying this strategy to QAOA is ineffective because of the QAOA circuit’s highly constrained structure. In particular, the limited expressiveness of the QAOA ansatz prevents it from representing a high-quality stabilizer state~\cite{cheng2022cliffordcircuitinitialisationvariational}.

\begin{figure}[t]
    \centering
    \includegraphics[width=0.9\columnwidth,trim={0cm 0cm 0cm 0cm},clip]{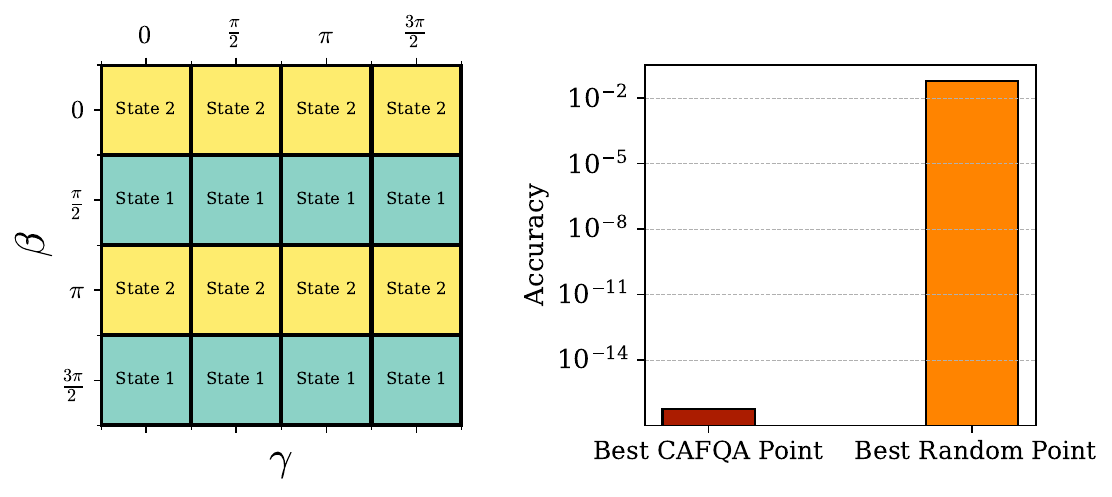}
    \caption{(Left): Heatmap of unique Clifford states found by CAFQA, with both states yielding expectation values $\approx 0$, indicating that only a few states are explored and resulting in low-quality solutions. (Right): Comparison of best energies, showing that a random guess (best of 50) can outperform CAFQA initialization for QAOA.}
    \label{fig:motivation}
\end{figure}

We illustrate this limitation in Fig. \ref{fig:motivation} using a weighted Max-Cut instance on a 10-node complete graph, a typical QAOA toy problem. For a QAOA circuit with depth \(p=1\), the CAFQA search space contains only  \((4)^{2p} = 16\) Clifford angle combinations.
We classically evaluate the expectation value (or `energy') of each candidate Clifford point. The resulting discrete heatmap clearly shows that the Clifford points lie in a low-quality region of the landscape. To contextualize this, we compare against the best initialization from 50 random starts and report the \emph{accuracy}, here defined as the closeness of the initialization energy to the true classical optimum. CAFQA performs significantly worse than random initialization because the restricted Clifford parameter space available to QAOA provides too few high-quality candidates. This highlights a fundamental limitation of directly applying Clifford-based initialization strategies to QAOA.
To overcome the limitations of Clifford-based QAOA initialization described above, we instead seek to leverage a fundamentally orthogonal QAOA strategy, which we describe next.

While increasing the repetitions theoretically enhances the expressivity of the QAOA circuit~\cite{farhi14}, deeper circuits suffer from amplified noise on current hardware, which quickly degrades state fidelity~\cite{murali2019noise,wang2020noise}. Even on early fault-tolerant systems, added depth is undesirable due to the increased number of rotational gates and corresponding resource overhead ~\cite{eft,dangwal2025variationalquantumalgorithmsera}.These challenges have motivated advances in ansatz design. Among them, Multi-Angle QAOA (ma-QAOA)~\cite{herrman2021maqaoa} introduces a more expressive QAOA circuit structure that allows independent rotation angles within each layer without increasing circuit depth.


Inspired by this approach, we find that relaxing the correlation between QAOA parameters substantially \emph{improves the circuit’s capacity to generate high-quality Clifford states} on classical compute, providing a pathway to effective and scalable QAOA initialization. Building on this insight, we propose a versatile initialization framework that leverages the relaxed-parameter ansatz together with a Clifford-based search strategy to efficiently identify near-ground-state solutions classically. To further mitigate the risk of initializing near non-convex regions, we introduce two strategies that select Clifford initialization points from diverse regions, enabling a multi-start optimization that promotes broader landscape exploration, typically avoids local minima, and increases the likelihood of high-quality solutions. Fig. \ref{fig:intro_landscape} provides a high-level overview of our approach. Unlike existing initialization methods tailored to specific problems, our method applies broadly to diverse combinatorial optimization tasks, yielding a scalable and versatile QAOA framework suitable for all quantum computers from today’s NISQ devices to tomorrow's fault-tolerant (FT) systems.


\textbf{\textit{The primary contributions and insights of our work are:}}
\begin{enumerate}[leftmargin=*]
    \item We develop \textbf{SPIQ}: a scalable, Clifford-driven initialization framework for QAOA. SPIQ overcomes the poor QAOA initialization with VQE-style Clifford techniques and generalizes effectively across diverse optimization problems, without requiring problem-specific heuristics.
    \item We leverage the expressive structure of ma-QAOA~\cite{herrman2021maqaoa}, demonstrating that beyond reducing circuit depth, it provides an effective framework for exploring and identifying high-quality Clifford states, with its expressivity naturally scaling with problem size and capturing richer structure in complex instances.
    \item We demonstrate that high-quality Clifford initialization prunes the QAOA solution space, eliminating a large fraction of non-optimal states.
    \item We introduce two strategies that select high-quality Clifford points from different regions of the landscape to seed a multi-start optimization that mitigates non-convexity and device noise, and increases the likelihood of reaching high-quality solutions.
    \item Our evaluations across diverse QUBO and PCBO instances (including real-world datasets) show that the proposed approach consistently outperforms existing warm-start methods and attains up to \underline{99.9\%} of the optimal solution, demonstrating strong robustness and generalization.
\end{enumerate}

\section{Background \& Motivation}
\label{sec:background_motivation}
\subsection{Combinatorial Optimization Problems} 

Combinatorial optimization seeks the best solution from a finite but exponentially large set of discrete choices and appears widely in logistics \cite{logistics_prob}, finance \cite{finance_prob}, and operations research \cite{operations_prob}. These problems are typically NP-hard, making exact classical solutions intractable at scale. We focus on QUBO and PCBO formulations. QUBO involves a quadratic unconstrained binary objective, while PCBO extends this to polynomial objectives with constraints, increasing expressivity but also complexity. Classical approaches include exact methods (e.g., branch-and-bound and relaxations in solvers such as CPLEX and Gurobi~\cite{cplex, gurobi}), which guarantee optimality but scale exponentially, and heuristics such as simulated annealing and genetic algorithms~\cite{boros2007local}, which provide approximate solutions for larger instances. PCBO problems are often transformed into QUBO via quadratization~\cite{boros2002quadratization, anthony2017quadratization}, enabling reuse of QUBO solvers at the cost of additional variables. As a result, practical performance for both classes relies heavily on heuristic and approximation methods~\cite{kochenberger2014unconstrained, glover2019qubo}.

\subsection{Quantum Approximate Optimization Algorithm (QAOA)}

The QAOA is a hybrid quantum-classical algorithm tailored for solving optimization problems. At its core, QAOA constructs a parameterized quantum circuit composed of alternating $p$ layers of problem-specific and mixing unitaries, each controlled by a set of classical parameters. 

In QAOA, an optimization problem is encoded into a cost Hamiltonian \(H_C\) and paired with a mixer Hamiltonian \(H_M\) to guide the variational evolution. The unitary operators \(U(\gamma, H_C) = e^{-i H_C \gamma}\) and \(U(\beta, H_M) = e^{-i H_M \beta}\), with variational parameters \(\gamma\) and \(\beta\), are applied alternately for \(p\) repetitions to evolve an initial quantum state towards low-energy configurations of \(H_C\). The circuit is initialized in the uniform superposition of all computational basis states, \(|s\rangle = \frac{1}{\sqrt{2^n}} \sum_{z} |z\rangle\). This construction is referred to as the traditional QAOA ansatz. QAOA relies critically on the choice of variational parameters $(\vec{\gamma}, \vec{\beta})$, which control the depth-$p$ quantum evolution. Optimizing these parameters is challenging due to the non-convexity and instance-specific structure of the cost landscape~\cite{wang18, farhi22, barak15, hastings19, marwaha21, chou22, lin16, farhi15, hadfield19, streif20}. These challenges arise, in part, from fundamental differences between QAOA and other variational algorithms, such as the VQE. While QAOA shares structural similarities with VQE, it targets discrete optimization over classical cost functions rather than ground-state estimation. Its layered, problem-specific ansatz constrains expressivity, making performance highly sensitive to parameter initialization. Increasing the circuit depth $p$ can enhance expressivity but also amplify noise, further complicating optimization. Together, these factors underscore the importance of principled initialization strategies to efficiently guide QAOA toward high-quality solutions. Addressing these challenges begins with problem formulation, as it directly impacts QAOA circuit design and parameter initialization. We consider instances in QUBO, PUBO, and PCBO forms, which encode objectives using binary variables and polynomial cost functions. A QUBO is defined as $\min x^T Q x$, where $x \in \{0,1\}^n$ and $Q \in \mathbb{R}^{n \times n}$ is symmetric. A PUBO extends this formulation to higher-order polynomial terms, $\min \sum_i a_i x_i + \sum_{i<j} b_{ij} x_i x_j + \sum_{i<j<k} c_{ijk} x_i x_j x_k + \cdots$. A PCBO further generalizes this setting by incorporating polynomial constraints.
\subsection{Parameter Relaxed Ansatz for QAOA}
The standard QAOA ansatz is minimally parameterized, typically using only $2p$ parameters. For increasing expressivity as the problem size grows, one could increase the number of classical parameters per QAOA layer. This is achieved by leveraging the multi-angle QAOA (ma-QAOA) \cite{herrman2021maqaoa}, which enhances the traditional structure by assigning individual parameters to each term in the cost and mixer operators, instead of using a single angle for each. Specifically:

\begin{align}
U(H_C,\vec{\gamma}_\ell)
&= e^{-i\sum_{a=1}^{m} H_{C,a}\gamma_{\ell,a}}
 = \prod_{a=1}^{m} e^{-iH_{C,a}\gamma_{\ell,a}}, \\
U(H_M,\vec{\beta}_\ell)
&= e^{-i\sum_{b=1}^{n} H_{M,b}\beta_{\ell,b}}
 = \prod_{b=1}^{n} e^{-iH_{M,b}\beta_{\ell,b}}.
\end{align}

\begin{figure}[t]
    \centering
    \includegraphics[width=\columnwidth,trim={1cm 14.5cm 14cm 4cm}]{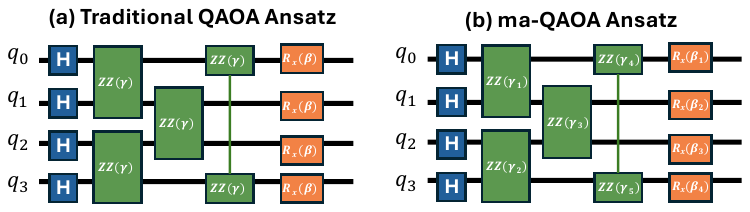}
    \caption{Structural differences between a $p=1$ (a) traditional QAOA ansatz and the (b) Multi-Angle QAOA ansatz}
    \label{fig:ansatz_diff}
\end{figure}

Here, $m$ is the number of clauses and $n$ is the number of qubits. The total number of parameters becomes $(m+n)*p$. The standard QAOA is a special case of ma-QAOA, where all parameters within a cost or mixer layer are identical. ma-QAOA expands the parameter space, allowing the exploration of more complex solution landscapes and improving optimization efficiency. The difference in ansatz structure between the traditional QAOA and ma-QAOA is illustrated in Fig. \ref{fig:ansatz_diff}.

\subsection{Clifford-Based Initialization in VQE}
Clifford gates map Pauli operators to Pauli operators under conjugation, enabling efficient classical simulation~\cite{gottesman1998heisenberg} and structured exploration of tractable regions of Hilbert space. This makes Clifford states appealing as initializations for variational algorithms. Prior work such as CAFQA~\cite{CAFQA_Ravi2022} leverages this to improve VQE by narrowing the optimization landscape and achieving high overlap with ground states. However, their effectiveness depends on ansatz expressivity. Unlike VQE, QAOA uses a fixed layered ansatz generated by the cost Hamiltonian $H_C$ and a non-commuting mixer $H_M$, restricting the reachable state space and complicating integration of pre-optimized Clifford states. As a result, Clifford-based initialization methods like CAFQA have largely remained limited to VQE and have not been applied to combinatorial optimization.

\subsection{Importance of Initialization in QAOA}
The success of the QAOA depends heavily on choosing the right starting values for its variables~\cite{Sack_2021,blekos2024review}. Without proper initialization, the algorithm often gets stuck in poor local minima or faces ``barren plateaus" where it cannot find the right direction to improve the solution~\cite{lee_2022}. To address this, researchers use specific techniques like \textbf{INTERP}, which builds complex circuits using patterns from simpler ones \cite{INTERP}, or \textbf{TQA}, which mimics a natural quantum annealing process \cite{Sack_2021}. However, these methods face significant limitations as they are often problem-specific and do not scale well, requiring costly custom adjustments as the complexity of the task changes. This lack of a general, scalable approach remains a major barrier to using QAOA for diverse, real-world applications.

\subsection{Red-QAOA}

As the problem size grows, larger circuits amplify noise and distort the optimization landscape. Red-QAOA~\cite{wang2024red} addresses this by using graph reduction to identify smaller, structurally equivalent graphs, allowing the parameter-search phase to run on shallower, less noise-sensitive circuits. The resulting parameters are then transferred back to the full graph as initial points. While producing encouraging results, Red-QAOA is fundamentally graph-centric, limiting it to problems with natural graph representations. Reformulating non-graph or higher-order problems is nontrivial. The method also incurs quantum resource overhead by requiring the execution of both reduced and full circuits. Additionally, landscape-preservation guarantees weaken for weighted graphs, where structured edge weights can degrade solution quality. These constraints highlight the need for scalable, problem-agnostic classical initialization strategies.

\subsection{Iterative VQA Optimization}

After SPIQ selects the initial points, a classical optimizer continuously updates the variational parameters over $[-\pi,\pi)$. Thus, the subsequent QAOA optimization is not restricted to Clifford angles.

\subsection{Multi-Start Optimization Strategies}
In complex, non-convex optimization landscapes such as those encountered in QAOA, the algorithm's performance is often highly sensitive to the initial parameter choice. A single optimization run, even from a well-chosen warm-start point, risks converging to a sharp or undesirable local minimum~\cite{Sack_2021, lee2024iterative}. A multi-start optimization strategy~\cite{rinnooy1987stochasticI} mitigates this risk by launching multiple independent runs from a diverse set of initial points, or seeds, and selecting the best result. In this work, we develop two strategies for selecting high-quality starting points to maximize exploration and improve convergence through multi-start.

\section{End-to-End Workflow} 
\begin{figure}[t]
    \centering
    \includegraphics[width=0.9\columnwidth,trim={4cm 9.1cm 6cm 3.65cm},clip]{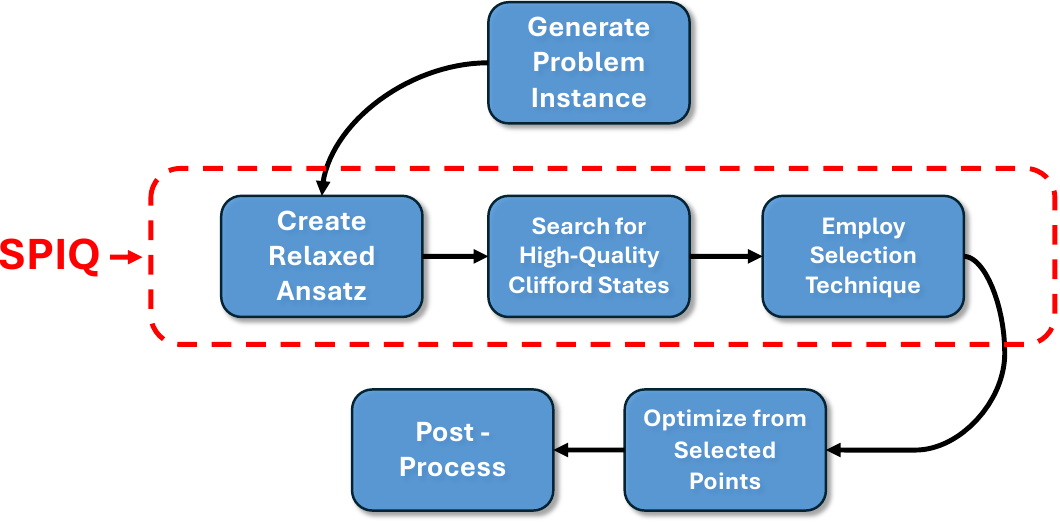}
    \caption{Overview of the QAOA + SPIQ workflow}
    \label{fig:illustrate_design}
\end{figure}

We propose a framework that leverages the expressive solution space enabled by the ma-QAOA ansatz to efficiently search for high-quality Clifford states as initial points. The following subsections describe each component of the framework in detail, and the framework is illustrated in Fig. \ref{fig:illustrate_design}. At a high level, the steps in the QAOA workflow are:
\begin{enumerate}
    \item \textbf{Generate Problem Instance:} Formulate the optimization problem as a QUBO or PCBO instance and map it to a Hamiltonian, where the ground state corresponds to the optimal solution of the original cost function.
    \item \textbf{Build ma-QAOA Ansatz:} Create a multi-angle ansatz intended to represent the problem Hamiltonian.
    \item \textbf{Search Clifford Space for Initial Parameters:} Perform a classical search for high-quality Clifford states.
    \item \textbf{Strategic Selection of Clifford Points:} Strategically choose Clifford initialization points from different regions to seed a multi-start optimization, improving exploration, avoiding local minima, and increasing the likelihood of high-quality solutions.
    \item \textbf{Optimize from Selected Points:} Initialize the QAOA ansatz with the chosen Clifford states and perform classical optimization of the variational parameters to iteratively minimize the cost function.
    \item \textbf{Post-Process:} Sample the final quantum state to extract the most probable bitstring, and the resulting output is interpreted as a solution for the optimization problem.
\end{enumerate}

Steps \textbf{1}, \textbf{5}, and \textbf{6} align with the standard QAOA workflow; further details appear in Section~\ref{sec:background_motivation}, where we outline the conventional QAOA process. Here, we present the core components of SPIQ.

\subsection{Relaxed Ansatz}

Once the cost Hamiltonian is constructed, it forms the foundation for building the QAOA ansatz, which governs the evolution of the quantum state. To enhance the ansatz’s expressiveness, we employ the ma-QAOA structure. We start by constructing a standard QAOA ansatz using Qiskit’s \texttt{QAOAAnsatz} API~\cite{qiskit2024}, which provides the traditional $2p$-parameter ansatz. We then relax the parameters $(\vec{\gamma}, \vec{\beta})$, allowing them to vary independently and increasing flexibility. This results in a total of $(m+n)*p$ parameters, where $m$ is the number of clauses and $n$ is the number of qubits. The ma-QAOA structure is illustrated in Fig. \ref{fig:ansatz_diff}(b).

\subsection{Clifford Search}
We leverage the ma-QAOA structure as it offers a broader solution space for exploring high-quality Clifford states. We note that this search is entirely classical since the execution of Clifford-only circuits only requires polynomial time~\cite{gottesman1998heisenberg}.  We then utilize a discrete search technique to explore Clifford parameters that minimize the expectation value of the quantum circuit with respect to the cost Hamiltonian. To perform this search, we leverage a gradient-free Genetic Algorithm (GA)~\cite{alam2020geneticalgorithmreviewsimplementations} optimizer to navigate the Clifford solution space.  While any discrete optimizer can be used, we choose GA since it offers substantial parallelism: each individual in a generation can be generated independently, making every generation fully parallelizable. The GA is implemented using the PyGad library \cite{pygad}. Since these parameters dictate the angles in the rotational gates of the ansatz, they can only take values that are integer multiples of \(\frac{\pi}{2}\), corresponding to rotations by \(0\), \(\frac{\pi}{2}\), \(\pi\), or \(\frac{3\pi}{2}\) radians. These specific angles ensure that the quantum operations remain within the Clifford group.

\subsection{Strategic Selection of Clifford Points}
Because the Clifford search is agnostic to the surrounding landscape, the lowest-energy Clifford state may still lie near a sharp local minimum or barren plateau, which can hinder QAOA tuning.  A multi-start optimization strategy~\cite{rinnooy1987stochasticI} mitigates this risk by launching several independent optimization runs from diverse seeds and retaining the best solution across them. The performance of the multi-start strategy is critically dependent on the method used to select these seeds. To address this, we introduce two complementary strategies for selecting high-quality, landscape-aware seeds that reliably strengthen the search from different regions of the landscape. 

Our first strategy is the \emph{Fixed-Interval Selection} method, which orders candidates by expectation value and picks points at regular intervals along this sorted list. This offers a direct way to highlight promising candidates while maintaining broad coverage. While this method ensures diversity in expectation value, it ignores the geometric structure of the parameter space, and the selected points may still be clustered in the same basin of attraction or reside in a stationary region, where a subsequent optimization algorithm cannot easily find a path for improvement~\cite{Larocca_2025}.
To address this, we introduce a second method, \emph{K-GAPS} (\underline{K}-means \underline{G}radient-\underline{A}ware \underline{P}oint \underline{S}election), that leverages K-means clustering and gradient-norm heuristics to identify effective Clifford states.

First, K-means clustering~\cite{kmean} is utilized to ensure spatial diversity. As an unsupervised algorithm, K-means partitions the candidate solutions into $k$ clusters based on \emph{distance} in the high-dimensional parameter space, ensuring that the selected seeds lie in structurally distinct regions. A key aspect of our approach is how we define \textit{distances} between seed points. Each coordinate of a seed corresponds to a VQA circuit parameter, which is a \emph{circular angle} in $[0,2\pi)$ (for Clifford states, ${0, \frac{\pi}{2}, \pi, \frac{3\pi}{2}}$). If we cluster raw angles directly, points like $0$ and $2\pi-\varepsilon$ appear far apart, even though they represent nearly identical states. To address this, we embed each angle $\theta$ onto the unit circle as $(\cos\theta, \sin\theta)$, capturing its periodicity. We then apply K-means to these 2D embeddings for every parameter, producing distances that meaningfully reflect the geometry of the VQA parameter space. Although each of these components (angle embedding, geometric distance, and clustering) can be considered separately, we combine them to effectively select diverse and representative seeds.

\begin{figure}[t]
    \centering
    \includegraphics[width=0.7\columnwidth,trim={1cm 10.9cm 6cm 3.8cm},clip]{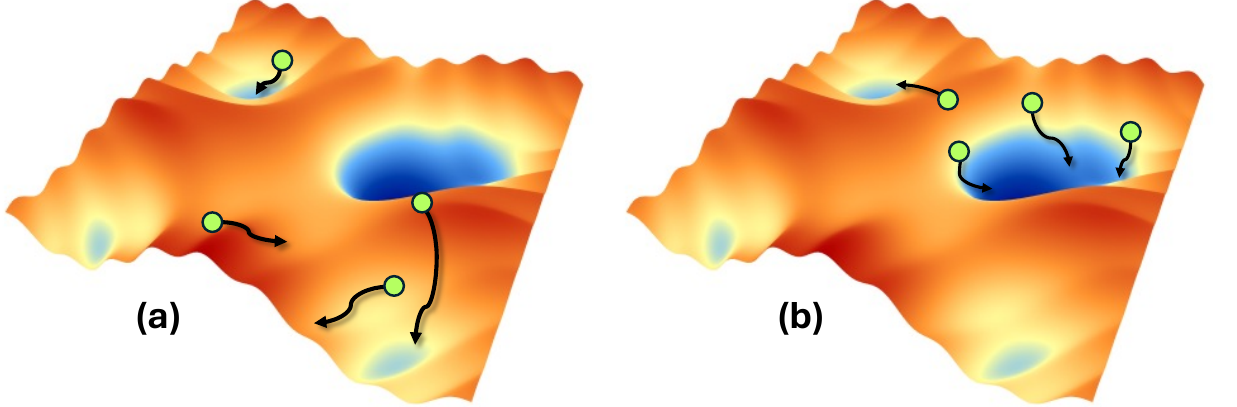}
    \caption{Illustration comparing two strategies for selecting points from high-quality SPIQ states: (a) naive selection may get stuck in local minima or barren plateaus, (b) choosing diverse points near steep curves helps avoid non-convex regions.}
    \label{fig:point_selection}
\end{figure}

Second, the gradient norm is used to filter for optimization potential. The gradient norm quantifies the steepness of the landscape at a point $\boldsymbol{\theta}$. For a function $f$ with $n$ parameters $\boldsymbol{\theta}=(\theta_1,\dots,\theta_n)$, the gradient norm is the $L_2$ norm~\cite{nocedal2006numerical}:

\begin{equation}
\|\nabla f(\boldsymbol{\theta})\|_2
=
\sqrt{\sum_{i=1}^{n}
\left(\frac{\partial f}{\partial\theta_i}\right)^2}.
\end{equation}
For a variational quantum circuit, partial derivatives are computed using the parameter-shift rule~\cite{para-shift}. For a single parameter $\theta_j$,

\begin{equation}
\frac{\partial f(\boldsymbol{\theta})}{\partial\theta_j}
=
\frac{1}{2}
\left[
f\!\left(\boldsymbol{\theta}+\frac{\pi}{2}\mathbf{e}_j\right)
-
f\!\left(\boldsymbol{\theta}-\frac{\pi}{2}\mathbf{e}_j\right)
\right],
\end{equation}

where $\mathbf{e}_j$ is the $j$th standard basis vector. Here, the parameter is shifted by $\pm\frac{\pi}{2}$. Since the candidate seeds are Clifford states, these shifts preserve the Clifford structure of the circuit, enabling rapid gradient evaluation via efficient stabilizer simulation~\cite{gottesman1998heisenberg}. A near-zero gradient norm indicates that a point already lies in a stationary region, such as a plateau or local minimum, making it a poor candidate for subsequent optimization. Fig.~\ref{fig:point_selection}(b) illustrates this selection process, which ensures that the selected seeds are both high-quality and spatially diverse while retaining strong potential for further variational improvement.

\section{Methodology}

\subsection{Benchmarks}
\label{sec:benchmarks}
\subsubsection{QUBO Benchmark} 

\label{subsubsec:qubo_benchmarks} 
\begin{figure}[t]
    \centering
    \includegraphics[width=0.7\columnwidth, trim={2.5cm 7.5cm 2.5cm 6.8cm}, clip]
    {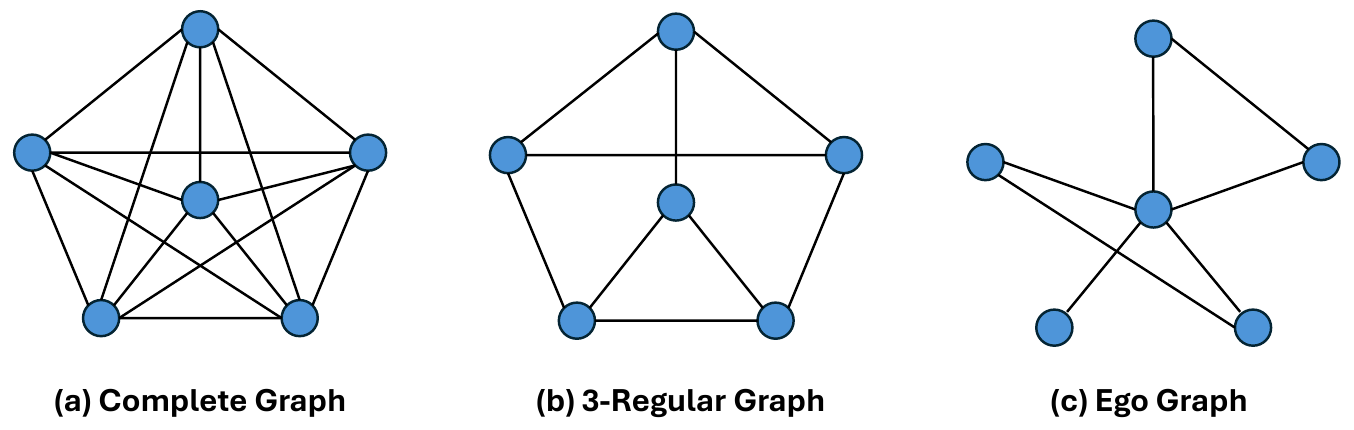}
    \caption{Illustration of graphs used for Max-Cut with 6 nodes.
    }
    \label{fig:illustrate_graphs}
\end{figure}

We select Max-Cut as the QUBO benchmark. The cost Hamiltonian for a graph \( G \) with edge set \( E(G) \) is typically expressed as
$H_{\text{C}} = \sum_{(i,j) \in E(G)} \frac{1}{2} w_{ij} \left( I - Z_i Z_j \right),$
where \( w_{ij} \) are the edge weights and \( Z_i \), \( Z_j \) are Pauli-\(Z\) operators acting on qubits corresponding to nodes \( i \) and \( j \). The solution is encoded in the ground state of this Hamiltonian. Max-Cut provides a suitable testbed for SPIQ, allowing exploration of diverse graph structures under consistent optimization settings. We evaluate performance on three graph types, illustrated in Fig. \ref{fig:illustrate_graphs}: Complete graphs, 3-Regular graphs, and Ego graphs. Complete graphs are densely connected, creating a highly entangled optimization landscape, while 3-Regular graphs are sparsely connected, with each node linked to three others, producing more localized interactions. Ego graphs, built around a central “ego” node and its neighbors, capture intermediate connectivity and local community structure. Weighted graph instances are generated by randomly assigning integer edge weights between 0 and 10. 

\subsubsection{PCBO Benchmarks}
\label{subsubsec:pcbo_benchmarks}
We evaluate PCBO performance on two representative benchmarks: the Knapsack problem and a medical higher-order interaction task~\cite{cancer_genome}. The Knapsack problem is formulated as a PUBO Hamiltonian using the Qiskit Optimization module~\cite{qiskit23}, with penalty terms enforcing the weight constraint, and random instances generated from item weights and values. The medical benchmark models gene-associated measurements, where each variable indicates inclusion in a selected subset and the objective captures both individual relevance and higher-order dependencies (pairwise and triple-wise interactions) via hypergraphs. The task is to select \(M\) out of \(N\) measurements to maximize relevance while minimizing redundancy, enforced through a Hamming-weight constraint with a Lagrange penalty.

\subsection{Evaluation Metrics}
\label{subsec:eval_metrics}

The primary metric used in this work is \emph{Accuracy}, which quantifies the quality of our initialization relative to the true ground state. We define accuracy as $\text{Accuracy} = \frac{E_\text{init}}{E_\text{opt}}$, where $E_\text{init}$ is the objective value of the initial state provided by different techniques, and $E_\text{opt}$ is the objective value of the optimal ground state after applying consistent sign conventions. A higher accuracy indicates that the initialization produces a state closer to the true optimum, likely serving as a more effective starting point for subsequent QAOA optimization.

In addition to accuracy, we introduce the \textit{Reduction Factor (RF)} to quantify the reduction in the number of solution states between initial points. Specifically, this metric captures how concentrated the quantum state becomes after initialization, by comparing the number of unique bitstrings sampled. We define
$\text{RF} = \frac{N_{\text{random}}}{N_{\text{init}}}$,
where \(N_{\text{random}}\) is the number of distinct bitstrings observed when sampling from a randomly initialized ma-QAOA ansatz, and \(N_{\text{init}}\) is the number of distinct bitstrings observed after applying SPIQ. All measurements are performed using the same total number of shots. A higher RF indicates that the initialized state is more sharply concentrated over a smaller portion of the solution space, enabling more focused and efficient exploration during optimization.

For large-scale problem instances where the true ground state energy is intractable to compute, we introduce \emph{Relative Improvement} to quantify the effectiveness of initialization, defined as $\text{Relative Improvement} = E_\text{init} / E_\text{rand}$, where $E_\text{init}$ is the objective value of the state obtained from our Clifford-initialized framework, and $E_\text{rand}$ is the objective value of a randomly initialized state. A higher relative improvement indicates that the Clifford initialization provides a better starting point for optimization, guiding the algorithm toward lower-energy regions even when the exact solution is unavailable. 

While no single metric fully captures QAOA performance, the chosen metrics effectively assess initialization quality, the scalability of finding high-value starting points, and potential post-initialization tuning. We note that final converged performance remains important, but in this work it is treated as a downstream consequence of improved initialization rather than a primary optimization objective.

\subsection{Baseline and Evaluation Settings}
\label{subsec:eval_settings}
We use two baselines: standard QAOA with random parameter initialization and Red-QAOA~\cite{wang2024red}, a warm-start method applicable only to graph problems like Max-Cut (Section \ref{subsec:comparision_prior_work}). In our GA search, we allocate a fixed evaluation budget corresponding to 48 hours of search across four independent starts. This tradeoff is well-motivated in our setting, as classical preprocessing is comparatively inexpensive compared to quantum evaluations, and investing effort upfront reduces the cost of downstream quantum optimization. The Clifford search and tuning evaluations were run on Perlmutter's~\cite {nersc_perlmutter} CPU node. We compare its performance against alternative techniques, including the ma-QAOA ansatz with random initialization, across varying experimental conditions. We design our ma-QAOA ansatz with two repeating layers, corresponding to a depth of $p=2$. To assess the performance of these methods, we first create a QAOA ansatz with an input Hamiltonian using the Qiskit library~\cite{qiskit2024} and then free the parameters in each layer to be independent to create the relaxed ma-QAOA ansatz. For noiseless simulations, we use Qiskit's \texttt{StatevectorSimulator} to obtain exact quantum states. For noise studies, we apply depolarizing noise channels~\cite{Fowler_2012} to gates, resets, and measurements. For our evaluation described in Subsection \ref{subsec:post_quantum_exploration},
we use Qiskit's COBYLA optimizer~\cite{Zhang_2023} to explore the solution space. Additionally, we showcase the top-performing points for multi-start optimization, selected from both the Fixed-Interval and K-GAPS methods. 
\section{Evaluation}
\subsection{Initialization Performance across Benchmarks}
\label{subsec:initialization_accuracy}
We first assess the performance of our initialization strategy across the benchmark combinatorial optimization problems detailed in Section \ref{sec:benchmarks}, as visualized in Figures~\ref {fig:cafqa_accuracy_maxcut}, \ref{fig:cafqa_accuracy_knapsack}, and~\ref{fig:cafqa_accuracy_teague}. We present the initialization accuracies that are computed relative to the ground state energy of the Hamiltonian associated with each combinatorial problem.

\subsubsection{Max-Cut Problem on Complete, 3-Regular, and Ego Graphs}
\label{subsubsec:eval_maxcut}
In this section, we evaluate the quality of the high-quality initialization point generated by SPIQ for the ma-QAOA ansatz applied to three distinct graph types in the Max-Cut problem. More information on Max-Cut and the types of graph benchmarks is discussed in Subsection \ref{subsubsec:qubo_benchmarks}.

\begin{figure}[t]
    \centering
    \includegraphics[width=\columnwidth,trim={0cm 0.75cm 0cm 0cm}]{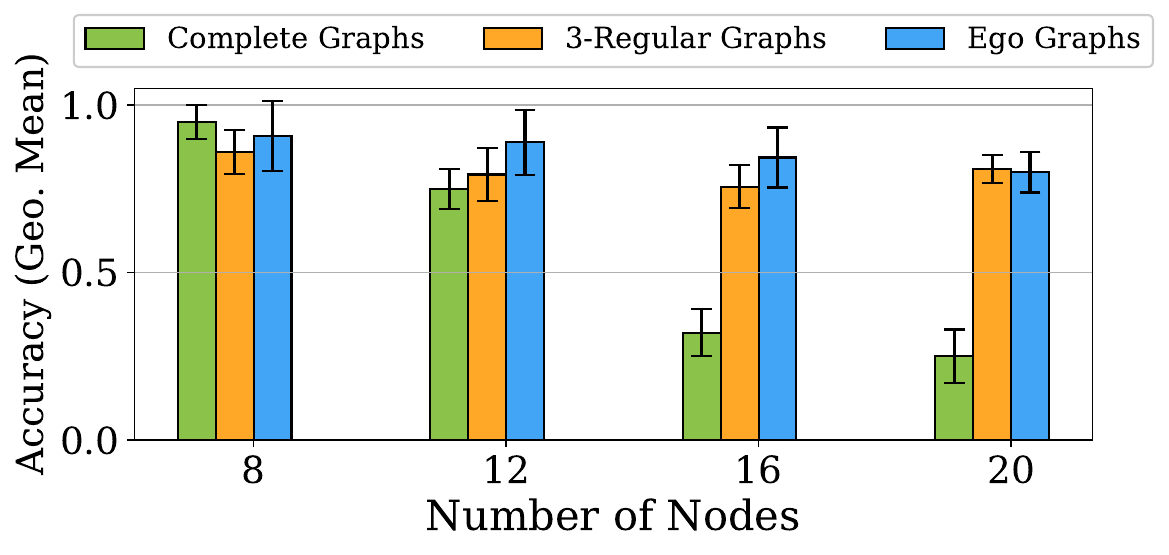}
    \caption{SPIQ Accuracy evaluated on Max-Cut problems for various weighted graphs.}
    \label{fig:cafqa_accuracy_maxcut}
\end{figure}

Fig. ~\ref{fig:cafqa_accuracy_maxcut} presents the initialization accuracies for each graph type across different numbers of qubits. We calculate the geometric mean over 10 weighted Max-Cut instances per graph type. In our weighted graphs, edges are assigned weights that modify the optimization objective to account for their relative importance rather than treating all edges equally.

We see that the Clifford initialization technique can find strong initialization points for diverse graph structures and sizes. Among the tested graphs, the Clifford initialization consistently achieves high proximity to the true ground state energy, suggesting it provides a strong starting point for optimization. The highest accuracy we achieve is 99.9\% for the 3-Regular and Ego graphs, which have fewer edges between the nodes than for complete graphs. The geometric mean accuracy decreases more sharply for the complete graphs as the number of nodes increases. This is expected as the edges between the nodes increase quadratically with the number of nodes, which makes the landscape extremely difficult to navigate. An increase in the number of edges directly translates to a larger set of cost-layer rotation parameters and mixer parameters, which enlarges the optimization space and makes the training process more challenging. We note that there are pre-processing techniques \cite{wang2024red} to further improve the quality of initialization that we don't explore in our work.

\subsubsection{Knapsack Problem}
Next, we assess initialization accuracy on a harder subclass of optimization problems, PCBO (see Subsection \ref{subsubsec:pcbo_benchmarks}), using the Knapsack problem as a representative example. In this section, we evaluate the impact of Clifford-based initialization on solving four representative Knapsack problem instances. We generate random 4, 9, 12, and 16-item problems, chosen specifically to align with the number of qubits used in other QUBO and PCBO benchmarks once the problems are translated into Hamiltonian form. This enables consistent comparisons across problem classes. As with the Max-Cut case, we set a 48-hour evaluation budget for the GA search.

\begin{figure}[t]
    \centering
    \includegraphics[width=\columnwidth,trim={0cm 0.65cm 0cm 0cm}]{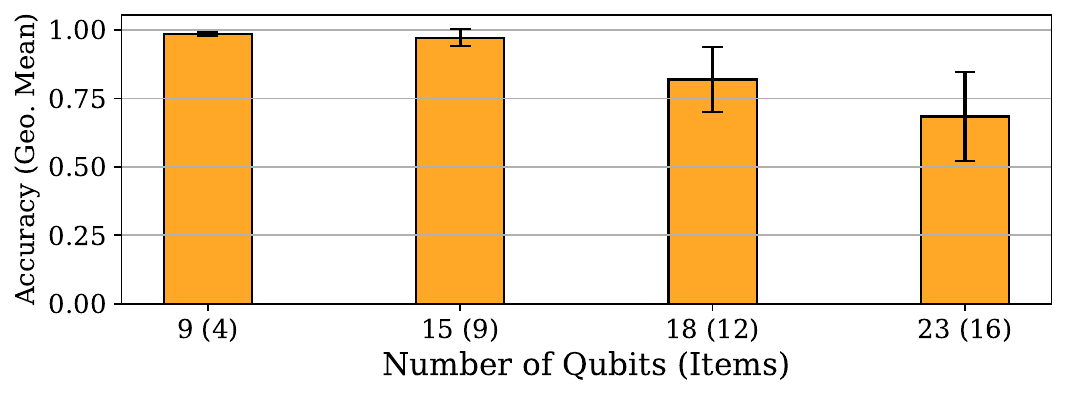}
    \caption{SPIQ Accuracy evaluated on Knapsack problems}
    \label{fig:cafqa_accuracy_knapsack}
\end{figure}

Fig. \ref{fig:cafqa_accuracy_knapsack} presents the quality of the initial solution from Clifford initialization across four spaced-out problem sizes. We again find that Clifford initialization yields strong overlap with the ground state, demonstrating its effectiveness even before any quantum optimization is performed. In the 4-item case (9 qubits), SPIQ recovers 99.9\% of the true energy on average. For the 9-item instance (15 qubits), it retains 98.2\% of the ground state energy. Although performance slightly decreases in the largest case (16 items, 23 qubits), recovering 70.2\% on average, this result demonstrates that the initialization strategy remains effective at producing low-energy configurations that are still significantly correlated with the ground state, even in high-dimensional, constrained Hamiltonians. The observed trend reflects the increasing difficulty of the optimization problem as the number of parameters grows, though longer runs of the Clifford search (or with more parallelized compute resources) can mitigate this, as discussed in {Subsection \ref{subsec:improve-cafqa-resource}}. 

\subsubsection{Medical High-Order Interaction Benchmark}

\begin{figure}[t]
    \centering
    \includegraphics[width=\columnwidth,trim={0cm 0.65cm 0cm 0cm}]{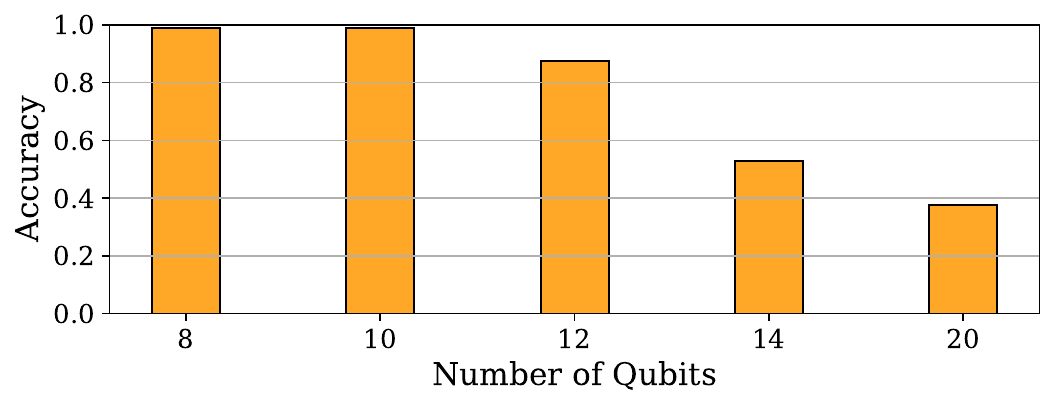}
    \caption{SPIQ evaluated on Medical High-Order Interaction}
    \label{fig:cafqa_accuracy_teague}
\end{figure}

In this section, we evaluate the Clifford initialization by applying it to a real-world medical biology challenge: selecting maximally informative yet non-redundant gene subsets from high-dimensional genomic data. We can frame the feature selection task as a PCBO problem. 
The objective function, whose formulation is described in Subsection \ref{subsubsec:pcbo_benchmarks}, 
creates a non-submodular optimization landscape that tests QAOA’s ability to handle highly constrained combinatorial problems. 
In our evaluations, we convert the PCBO object, which is represented as a hypergraph, into an Ising Hamiltonian formulation with the process described in Subsection \ref{subsubsec:pcbo_benchmarks}. In our study, the problems are subject to the constraint of choosing exactly 4 features out of the $N$ total features in the datasets. The benefits of Clifford-based initialization on this problem are illustrated in Fig. \ref{fig:cafqa_accuracy_teague}. The best-performing SPIQ-initialized point achieves 94\% accuracy, indicating close recovery of the ground state. These results demonstrate the value of SPIQ in steering the optimizer toward high-fidelity solutions, even as the problem complexity increases. Once again, we note that our current estimates are based on a maximum of 48 hours of runtime, but they can be improved with increased memory or compute resources, extended execution time, improved search heuristics, more efficient parallelization, or targeted exploration beyond Clifford space, as discussed in section \ref{subsec:improve-cafqa-resource}.

\subsection{Scaling to Larger Problem Instances}

\begin{figure}[t]
    \centering
    \includegraphics[width=\columnwidth,trim={0cm 0.65cm 0cm 0cm}]{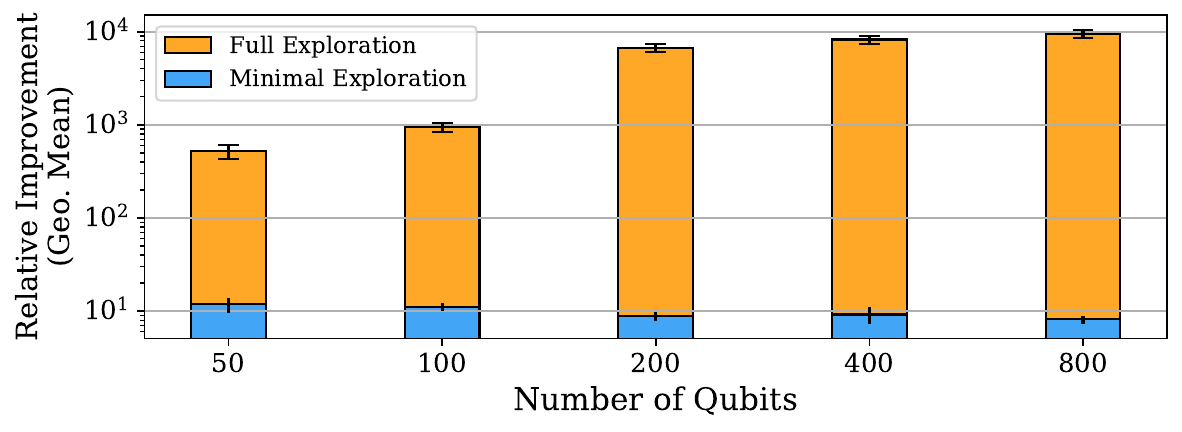}
    \caption{Relative Improvement evaluated on large 3-Regular Max-Cut instances.}
    \label{fig:cafqa_accuracy_large_maxcut}
\end{figure}

To assess the scalability of SPIQ, we apply it to larger problem instances. Specifically, we evaluate performance on Max-Cut problems defined over 10 instances of 3-regular graphs with sizes ranging from 50 to 800 qubits, representing a challenging yet structured class of optimization problems. To quantify the effectiveness of initialization at this scale, we compute the relative improvement, defined as the ratio between the energy of our Clifford-initialized state and that of a randomly initialized state. As shown in Fig. \ref{fig:cafqa_accuracy_large_maxcut}, our method achieves significantly better starting points, with improvements reaching up to 10,000× over random initialization. We report results for two exploration strategies: `Full Exploration' shows the best Clifford state identified after running the search up to the 48-hour maximum wall time, while `Minimal Exploration' indicates the first state found among all generations within the same wall-time limit, highlighting that even limited exploration can produce high-quality states at scales beyond the reach of other classically simulable initialization techniques. These results underscore the importance of informed initialization for steering the optimizer toward low-energy regions, even in large problems. Large initial gains reduce the quantum optimization burden, benefiting runtime-constrained or low-depth settings. Since the exact ground-state energy is intractable for these cases, this metric serves as a practical proxy for initialization quality.

\subsection{Reduction in Solution Space with Initialization}

\begin{figure}[t]
    \centering
    \includegraphics[width=\columnwidth,trim={0cm 0.65cm 0cm 0cm}]{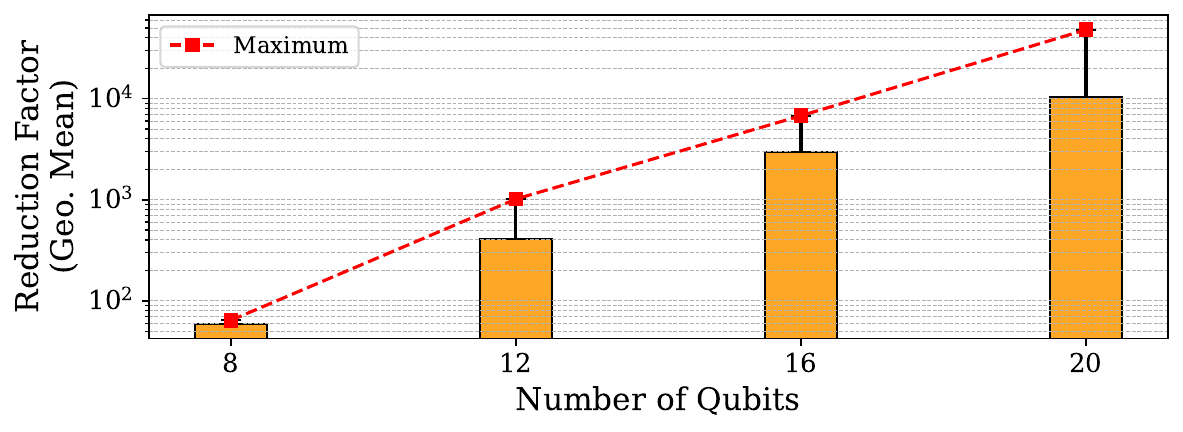}
    \caption{Factor of Reduction in Number of Initial Solution States in 3-Regular Max-Cut.}
    \label{fig:statevector_reduction_k_reg}
\end{figure}

\begin{figure}[t]
    \centering
    \includegraphics[width=\columnwidth,trim={0cm 0.65cm 0cm 0cm}]{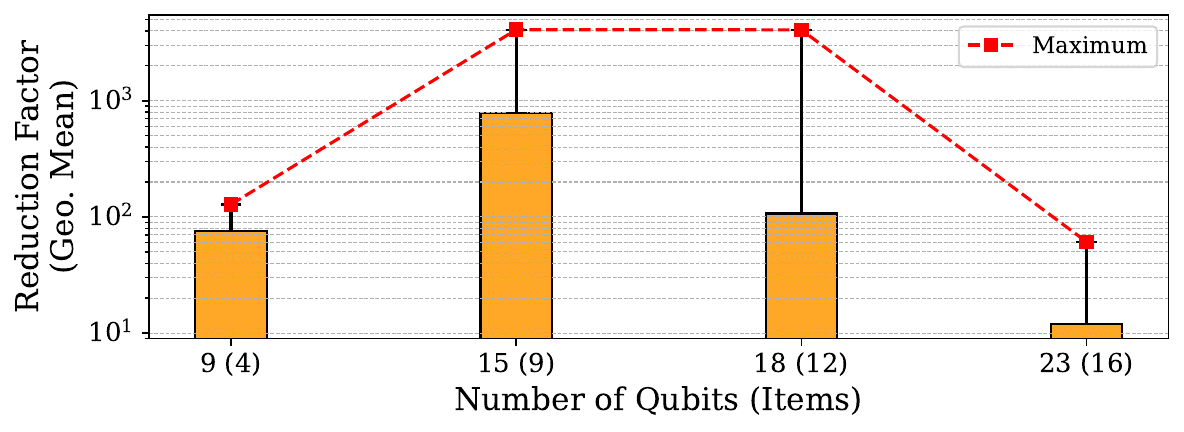}
    \caption{Factor of Reduction in Number of Initial Solution States in the Knapsack Problem.}
    \label{fig:statevector_reduction_knapsack}
\end{figure}

To evaluate how effectively our initialization method narrows the solution search space, we compare the number of solution states explored by QAOA with SPIQ against random initialization. We perform this analysis on both the Knapsack problem and 3-regular Max-Cut instances across several medium-sized problem sizes.  To accurately estimate the distribution over solution states, we sample the initialized QAOA states 100 million times. As shown in Fig. \ref{fig:statevector_reduction_k_reg}, our initialization method achieves a maximum reduction of 48,132× and an average reduction of 10,448× in the number of solution states compared to random initialization on 3-regular Max-Cut instances. Similarly, in Fig. \ref{fig:statevector_reduction_knapsack}, we observe strong reductions on the Knapsack instances, with a maximum reduction of 4,095× and an average reduction of 789× across the qubit sizes evaluated. Despite the added complexity of non-binary constraints and rugged energy landscapes, the initialization consistently concentrates amplitude into structured, low-energy regions of the solution space. Although the cause of the variability in Fig. ~\ref{fig:statevector_reduction_knapsack} is not fully understood, these results collectively show that SPIQ allows QAOA to start from compact, high-quality regions across a wide range of optimization problems.

\subsection{Comparison Against Prior Work}
\label{subsec:comparision_prior_work}

We also compare the performance of SPIQ against prior warm-start techniques. Previous works have proposed problem-specific initialization methods to improve the convergence and solution quality of quantum optimization algorithms. Among these, Red-QAOA~\cite{wang2024red} is one of the leading warm-start approaches and serves as the primary baseline for our evaluation. We evaluate the performance of Red-QAOA against SPIQ on both unweighted and weighted instances of Complete, 3-Regular, and Ego graphs. For each graph type, we select the best-performing initialization across 50 random restarts over 10 graph instances in a noiseless setting.

\begin{figure}[t]
    \centering
    \includegraphics[width=\columnwidth,trim={0cm 0.65cm 0cm 0cm}]{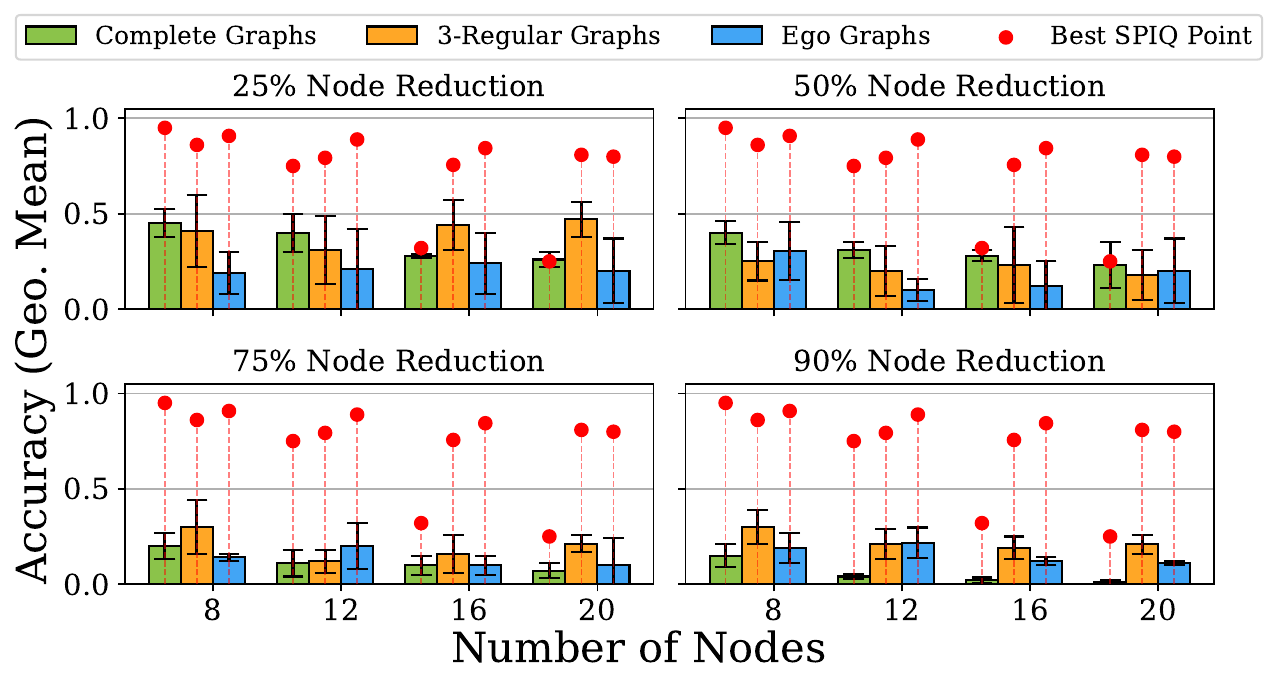}
    \caption{Performance of SPIQ (points) against Red-QAOA (bars) on weighted Max-Cut instances.}
    \label{fig:red-qaoa_weighted}
\end{figure}

\begin{figure}[t]
    \centering
    \includegraphics[width=\columnwidth,trim={0cm 0.65cm 0cm 0cm}]{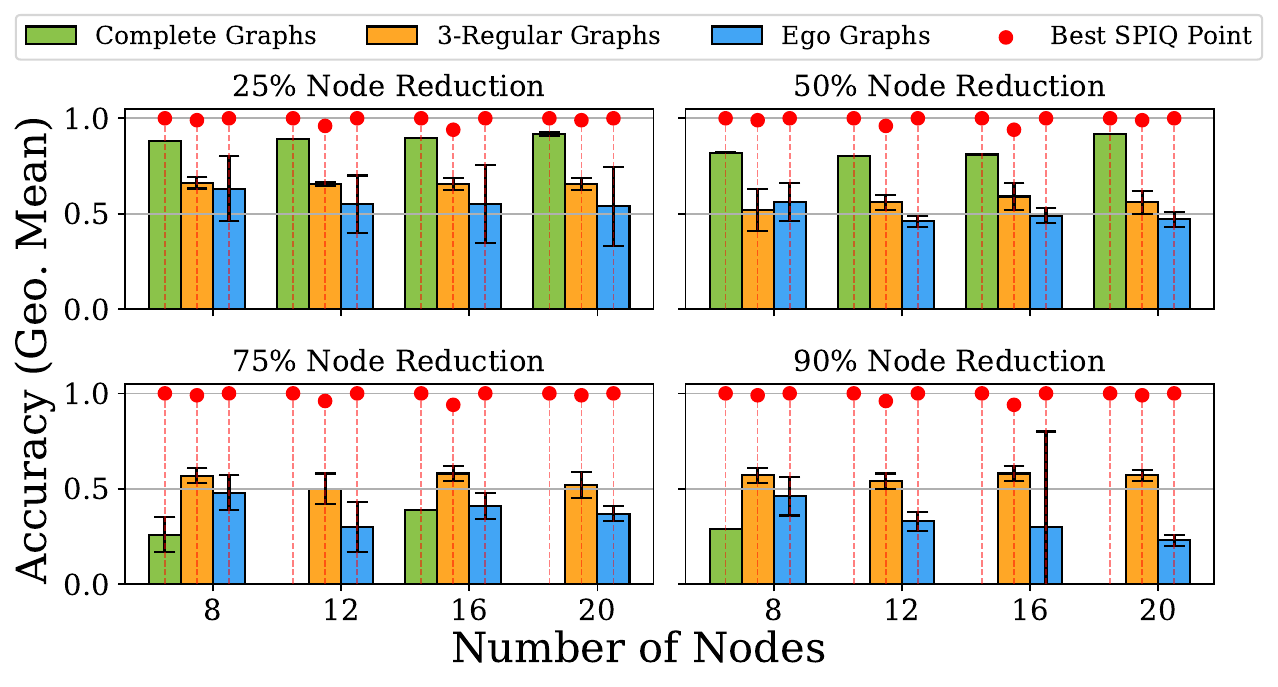}
    \caption{Performance of SPIQ (points) against Red-QAOA (bars) on unweighted Max-Cut instances.}
    \label{fig:red-qaoa_unweighted}   
\end{figure}

As shown in Fig. \ref{fig:red-qaoa_weighted}, we report results for weighted graph instances while varying the \emph{node reduction factor} in Red-QAOA. The node reduction factor quantifies the degree to which the graph’s size (in terms of nodes) decreases through the reduction process, directly influencing the feasibility of executing the resulting circuit on a smaller QPU. We observe that SPIQ consistently outperforms Red-QAOA across all graph types and node reduction factors. Specifically, while Red-QAOA achieves a maximum accuracy of approximately 60\%, SPIQ reaches 99.9\% accuracy, as discussed in greater detail in Subsection \ref{subsubsec:eval_maxcut}. We further compare the performance of SPIQ on unweighted graph instances, as originally evaluated in the Red-QAOA study, illustrated in Fig. \ref{fig:red-qaoa_unweighted}. Although Red-QAOA performs better on unweighted than on weighted graphs, SPIQ still achieves significantly higher accuracy across all configurations. Finally, we note a key limitation of graph-specific warm-start methods like Red-QAOA. These rely on reduced graph representations to preserve the QAOA landscape, but for PCBO problems such as Knapsack, which lack a natural graph structure, any graph encoding is artificial and non-canonical. This doesn't guarantee that the landscape is preserved, limiting Red-QAOA’s applicability beyond graph-based problems. Overall, our results demonstrate that while Red-QAOA remains effective within its intended problem class, its reliance on graph-reducible formulations constrains its generality. In contrast, SPIQ exhibits robust performance across diverse problem structures.

\subsection{Post-Initialization QAOA Exploration}
\label{subsec:post_quantum_exploration}

\begin{figure*}[t]
    \centering
    \includegraphics[width=\textwidth,trim={0.5 0.9cm 0.9cm 1cm 0cm}]{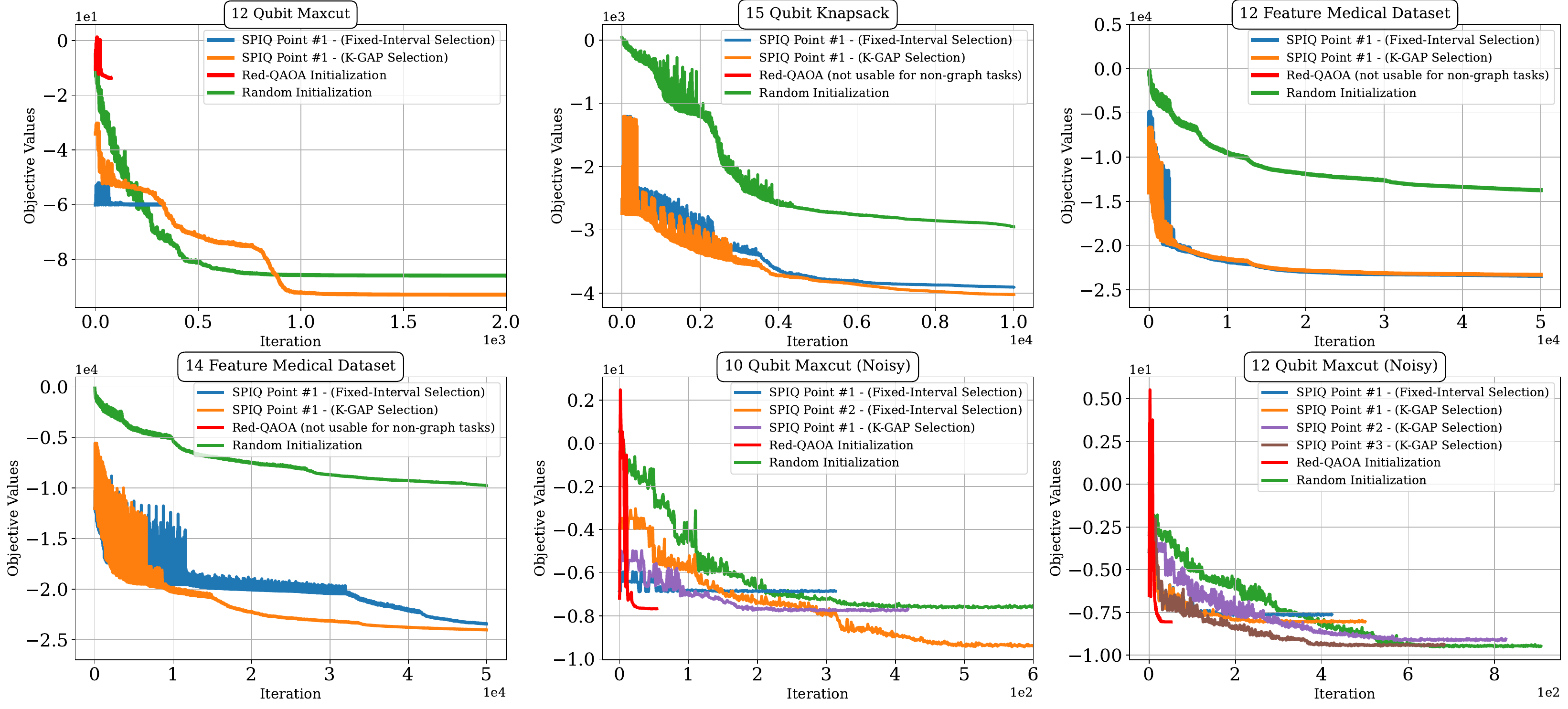}
    \caption{Effect of initialization choice on tuning across diverse problem instances.}
    \label{fig:multi-start_subplots}
\end{figure*}

A key factor in QAOA performance is effective exploration of the parameter space after initialization. Near- and intermediate-term devices are prone to noise, and rugged, plateau-filled landscapes can stall optimization~\cite{Larocca_2025}. We compare SPIQ-initialized trajectories against Red-QAOA and random starts across representative instances, as shown in Fig.~\ref{fig:multi-start_subplots}. To ensure diverse coverage, we use K-GAPS and Fixed-Interval to extract candidate points from SPIQ. On a 12-qubit weighted Max-Cut instance, K-GAPS yields the best-performing initialization, while Red-QAOA quickly plateaus at a low-quality solution. Even the best point selected by SPIQ’s Fixed-Interval method eventually stalls, underscoring the need for more advanced, landscape-adaptive SPIQ selection strategies. On constrained problems such as Knapsack and the medical dataset, SPIQ initializations consistently outperform random starts, demonstrating robustness across problem classes.   We also note that Red-QAOA is limited to graph-structured problems and is therefore not applicable to our non-graph benchmarks (e.g., Knapsack and higher-order interaction models), as graph reformulation is nontrivial and adds additional modeling overhead beyond the scope of this work.
We further evaluate a noisy regime using a depolarizing channel (single-qubit error of $10^{-4}$, two-qubit error of $10^{-3}$) on unweighted Max-Cut. In the 10-qubit case, Fixed-Interval avoids non-convex regions that trap other initializations. In the 12-qubit case, both SPIQ and random initialization reach similar final solutions, but K-GAPS accelerates convergence, indicating improved proximity to high-quality regions. Overall, these results highlight the importance of multi-start initialization: in noisy settings, selecting diverse high-quality starting points improves the likelihood of efficient convergence. While this work does not focus on quantum landscape exploration, which depends on multiple factors including problem structure, optimizer design, and device noise, Clifford-based initialization remains essential for driving towards high-quality solutions. While post-initialization evaluations are simulated under a depolarizing noise model and do not reflect execution on real quantum hardware, SPIQ provides starting points that reduce susceptibility to noise and barren plateaus, enabling faster convergence. Additionally, reducing repeated quantum executions offers substantial economic savings, as prior studies~\cite{gu2021adaptive,hou2026treevqa} show VQA on cloud platforms become prohibitively costly even for modest problem sizes.

\section{Discussion}
\paragraph{Initialization Beyond Max-Cut-Specific Approaches}
Prior QAOA initialization methods~\cite{wang2024red, wilkie2024angle} have focused mainly on Max-Cut, using instance-specific heuristics to identify promising regions of parameter space. While effective, they incur significant classical overhead and generalize poorly to larger or weighted problems, limiting applicability beyond Max-Cut-like instances. SPIQ addresses this by extending initialization to a broader class of combinatorial problems, including weighted and practically relevant formulations.
\paragraph{Multi-Start for Efficient Exploration}\label{subsec:multi-start}
VQA involves landscapes with many local optima~\cite{Larocca_2025}, making initialization critical. We use a multi-start approach with two selection methods to sample diverse regions of parameter space, reducing reliance on a single starting point. K-GAPS accounts for landscape steepness but is not consistently superior to Fixed-Interval, reflecting a problem-dependent exploration–exploitation tradeoff. Nonetheless, it promotes spatial diversity and identifies degenerate optima. Our results show multi-start strategies shape landscape coverage, with optimizers favoring different regions (Fig.~\ref{fig:multi-start_subplots}).
\paragraph{Classical Search and Resource Scaling}\label{subsec:improve-cafqa-resource}
Our Clifford search uses a GA optimizer to balance global exploration and robustness to noisy evaluations. Increasing classical resources enables deeper and broader searches, making optimizer choice and resource allocation important for larger problems.
\paragraph{Relevance in Fault-Tolerant Devices}
Although VQAs are often associated with NISQ devices, QAOA remains relevant for intermediate and early fault-tolerant (EFT) regimes, where efficient quantum resource usage is still critical~\cite{dangwal2025variationalquantumalgorithmsera,eft}. SPIQ operates at a high abstraction level and is agnostic to the underlying gate set, enabling the same initialization benefits on fault-tolerant systems.

\section{Conclusion}

We introduce an initialization framework that leverages the relaxed ma-QAOA ansatz to efficiently identify high-quality Clifford states as QAOA starting points, improving solution quality and reducing resource costs. We demonstrate its effectiveness across a range of problems, including Max-Cut, Knapsack, and real-world medical higher-order interaction datasets, consistently achieving higher overlap with the ground-state energy. These results highlight the role of improved initialization in making QAOA more practical for larger combinatorial optimization workloads.
\section*{Acknowledgement}
We thank Teague Tomesh and Frederic Chong for helpful and insightful discussions. 
This material is based upon work supported by the U.S. Department of Energy, Office of Science, Office of   Scientific Computing Research, Accelerated Research in Quantum Computing under Award Number DE-SC0025633. This research used resources of the National Energy Research Scientific Computing Center, a DOE Office of Science User Facility supported by the Office of Science of the U.S. Department of Energy under Contract No. DE-AC02-05CH11231 using NERSC award ASCR-ERCAP0033197 and NERSC DDR-ERCAP0035341.


\bibliographystyle{IEEEtranS}
\bibliography{ref,ref_yw}

\end{document}